\documentclass[apj]{emulateapj}

\usepackage{graphics}
\usepackage{epsfig}
\usepackage{apjfonts}
\shorttitle{Narrow double-peaked emission lines of SDSS J1316+1753}
\shortauthors{D. Xu \& S. Komossa}

\newcommand{\kms}{km s$^{-1}$}

\newcommand{\oiii}{[\ion{O}{3}]}
\newcommand{\oii}{[\ion{O}{2}]}

\newcommand{\caii}{\ion{Ca}{2}}
\newcommand{\nai}{\ion{Na}{1}}

\newcommand{\sii}{[\ion{S}{2}]}
\newcommand{\nii}{[\ion{N}{2}]}
\newcommand{\neiii}{[\ion{Ne}{3}]}

\newcommand{\ariii}{[\ion{Ar}{3}]}

\newcommand{\msun}{${\rm M}_{\odot}$}

\slugcomment{Submitted 2009 August 5; accepted 2009 September 11}
\journalinfo{ApJ Letters, 705, L20-L24, 2009 (November 1 issue)}

\begin{document}

\title{Narrow double-peaked emission lines of SDSS J131642.90+175332.5: 
signature of a single or a binary AGN in a merger, jet-cloud interaction,
or unusual narrow-line region geometry}

\author{Dawei Xu\altaffilmark{1} and S. Komossa\altaffilmark{2}}

\affil{$^1$National Astronomical Observatories, Chinese Academy of Sciences,
Beijing 100012, China; dwxu@bao.ac.cn\\
$^2$Max-Planck-Institut f\"ur extraterrestrische Physik,
Postfach 1312, 85741 Garching, Germany; skomossa@mpe.mpg.de}

\begin{abstract}

We present an analysis of the active galaxy SDSS\,J131642.90+175332.5, 
which is remarkable because all of its narrow emission lines are 
double-peaked, and because it additionally shows an extra broad 
component (FHWM $\sim$\,1400 km\,s$^{-1}$) in most of its forbidden lines,  
peaking in between the two narrow systems.
The peaks of the two narrow systems are separated by 
400--500 \kms\ in velocity space. 
The spectral characteristics of
double-peaked \oiii\ emission have 
previously been interpreted as a signature of dual or binary active 
galactic nuclei (AGNs), among other models. 
In the context of the binary scenario, 
SDSS\,J131642.90+175332.5 is a particularly good candidate
because not just one line but all of its emission lines are double-peaked.
However, we also discuss a number of other scenarios
which can potentially account for double-peaked narrow emission lines,
including projection effects, 
a two-sided outflow, 
jet-cloud interactions, 
special narrow-line region (NLR) geometries (disks, bars, or inner spirals), 
and a galaxy merger
with only {\em one} AGN illuminating {\em two} NLRs.
We argue that the similarity of the emission-line ratios in both systems, and
the presence of the very unusual broad component at intermediate velocity, 
makes a close pair of unrelated AGNs unlikely, and rather 
argues for processes in a single galaxy or merger.
We describe future observations which can distinguish between these
remaining possibilities.

\end{abstract}


\section{Introduction}

According to hierarchical models of galaxy formation,
galaxies will merge frequently with each other,
forming supermassive binary black holes (SMBBHs) at
their centers. The merging of the two SMBHs 
would proceed in three stages  (Begelman et al. 1980).
After an initial phase 
of merging of the galaxy cores by dynamical friction,
the two SMBHs would form a bound pair. The duration of this second phase
is still uncertain and depends on the efficiency of gas- or stellar-dynamical
processes to aid shrinking the binary orbit and prevent the SMBHs from stalling
(e.g., Merritt \& Milosavljevi{\'c} 2005).
Once they reach a separation of about 0.01 pc, emission of 
gravitational waves will
lead to the coalescence of the two black holes within a relatively short
time interval (e.g., Sathyaprakash \& Schutz 2009).
Measuring the frequency of wide and close pairs of SMBHs, and their properties,
is therefore important not only in the context of understanding 
galaxy mergers and
mechanisms of AGN fueling in dependence of the merger phase. It also allows
us to constrain timescales of SMBH
merging and coalescence, and to estimate
the fraction of gravitational wave sources detectable with the
{\sl LISA} mission.

The search for binary AGNs and SMBBHs has therefore received great attention,
yet few definite cases are known.
In recent years, a few pairs of accreting SMBHs have been found at the
centers of single galaxies based on spatially
resolved X-ray, radio, and optical imaging spectroscopy,
namely NGC 6240 (Komossa et al. 2003),
J0402+379 (Rodriguez et al. 2006; Morganti et al. 2009a), and
COSMOS J100043.15+020637.2 (Comerford et al. 2009b).
In addition, a number of candidates for very close pairs of SMBHs
have been reported, typically based on semi-periodic
or other structures in radio jets
and in multi-wavelength light curves (e.g., Lobanov \& Roland 2005;
Valtonen et al. 2008; see Komossa 2006 for a review).
In particular, a few galaxies with 
double-peaked \oiii, and perhaps H$\beta$, emission lines 
have been presented as candidate binary AGNs, each AGN with
its own narrow-line region (NLR; Zhou et al. 2004; Gerke et al. 2007; 
Comerford et al. 2009a).
Double-peaked \oiii\ emission has also
occasionally been observed in some nearby AGNs since
the 1980s (e.g., Heckman et al. 1981; 
Whittle et al. 1988; Veilleux 1991a, 1991b),
among them some that have all lines double-peaked.
At that time, processes linked to radio(jet) activity
had been suggested to explain these line profiles.
Narrow-line splitting is also evident in  
a few compact radio sources (e.g., Holt et al. 2008).

In a search for peculiar
emission-line AGNs in the Sloan Digital Sky Survey Data
Release 7 (SDSS DR7; Abazajian et al. 2009), we have
found SDSS J131642.90+175332.5 (SDSS J1316+1753
hereafter), which displays a rich system of narrow emission lines,
all of which are double-peaked.
In this Letter, we discuss this and other
remarkable properties of this galaxy.
We also critically examine a number of scenarios that can
potentially explain the observed features  
including a close pair of unrelated AGNs,
biconical outflows, jet-cloud interactions, 
special NLR geometries, 
and a binary or single AGN illuminating two NLRs
in a galaxy merger.
A cosmology with $H_{\rm 0}$=70 km\,s$^{-1}$\,Mpc$^{-1}$, $\Omega_{\rm M}$=0.3,
and $\Omega_{\rm \Lambda}$=0.7 is used throughout.

\section{Spectral analysis}

\begin{figure*}
\begin{center}
    \includegraphics[width=115mm]{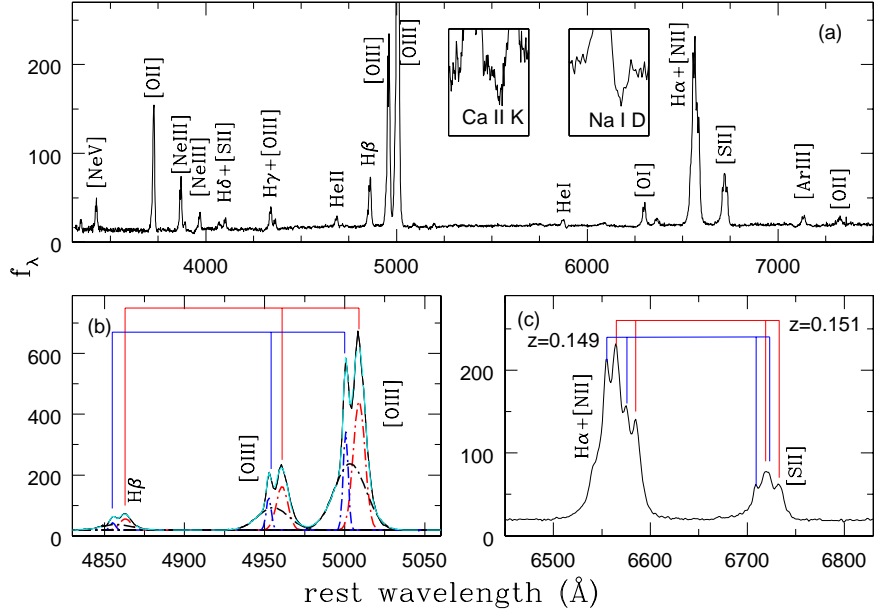}
\end{center}
\caption{
SDSS spectrum of SDSS J1316+1753.
The upper panel shows the full spectrum, with emission
lines labeled,
and the insets show a zoom on the absorption features.
The lower panel zooms into the [O\,III]-H$\beta$ and
the H$\alpha$-[N\,II]-[S\,II] regime, and shows the two sets of
narrow emission lines which are separated by 400--500\,\kms\
in velocity space.
The red system is marked in red and the blue system in blue.
The dot-dashed lines in the lower left panel
represent our Gaussian fitting with two narrow
lines (the red and blue system) and one intermediate broad component.
The flux density f$_\lambda$ is given in units of
10$^{-17}$~erg~s$^{-1}$~cm$^{-2}$~\AA$^{-1}$.
}
\end{figure*}

\begin{deluxetable*}{llllllllllllll}
\tabcolsep 1.0mm
\tablecolumns{14}
\tablewidth{0pt}
\tablecaption{Emission-line properties of SDSS\,J1316+1753}
\tablehead{
\colhead{Property}  &\colhead{[NeV]}        & \colhead{${\rm [OII]}$} & 
\colhead{${\rm [NeIII]}$} & 
\colhead{${\rm H\gamma}$}     & \colhead{${\rm [OIII]}$} & 
\colhead{${\rm HeII}$}    & \colhead{${\rm H\beta}$}     & 
\colhead{${\rm [OIII]}$}  & \colhead{${\rm [OI]}$}   & 
\colhead{${\rm H\alpha}$}     &\colhead{${\rm [NII]}$}   &
\colhead{${\rm [SII]}$}   & 
\colhead{${\rm [ArIII]}$} \\
\colhead{}  &\colhead{3426}           & \colhead{${\rm 3727}$} &
\colhead{${\rm 3869}$} 
&
\colhead{}     & \colhead{${\rm 4363}$} &
\colhead{${\rm 4686}$}    & \colhead{}     &
\colhead{${\rm 5007}$}  & \colhead{${\rm 6300}$}   &
\colhead{}     &\colhead{${\rm 6583}$}   &
\colhead{${\rm 6725}$} &
\colhead{${\rm 7136}$} 
}
\startdata
{\bf Blue system}  & & & & & & & & & & & & &   \\
Line ratio & 0.4  & 1.3  & 0.9  
& 0.3 
& 0.2  & 0.2  & 1.0\tablenotemark{a} & 12.4  
& 0.6  & 4.5  & 2.8  & 1.7\tablenotemark{b} & 0.4  \\
$z$  & 0.1489   & 0.1492  & 0.1490  
& 0.1491\tablenotemark{c} & 0.1491\tablenotemark{c}
& 0.1491 & 0.1491    &   0.1491   &   0.1492 
&   0.1491\tablenotemark{c}  &   0.1492\tablenotemark{c}  
&   0.1492     &   0.1491  \\
FWHM\tablenotemark{d}&140 &200 &140 
&180\tablenotemark{c}
&140\tablenotemark{c} &180\tablenotemark{c}
&180 &140 &170 &180\tablenotemark{c} &170\tablenotemark{c} &170 &170 \\
{\bf Red system}   & & & & & & & & & & & & &  \\
Line ratio & 0.7  &  1.9  & 1.1  
& 0.4  &  0.2  &  0.2   & 1.0\tablenotemark{a} 
& 13.9  & 0.4  & 4.0   & 2.5  & 2.1\tablenotemark{b} & 0.3  \\
$z$ &  0.1509  & 0.1509  & 0.1509  
& 0.1510\tablenotemark{c} &0.1510\tablenotemark{c}  
& 0.1511  &   0.1510 &   0.1510  & 0.1510 
& 0.1510\tablenotemark{c} & 0.1509\tablenotemark{c} & 0.1509  
&   0.1510  \\
FWHM\tablenotemark{d} &630 &430 &500 
&430\tablenotemark{c} 
&480\tablenotemark{c} &430\tablenotemark{c}
&430  &480 &290 
&430\tablenotemark{c} &400\tablenotemark{c}
&400  &450 \\
{\bf Broad system}   & & & & & & & & & & & & &  \\
Line ratio & 0.3  & 2.1  & 0.7  
&0.5  &0.1  &0.2   &1.0\tablenotemark{a}
&10.4  &0.7  &5.1  &3.0  &1.9\tablenotemark{b} &0.2  \\
$z$  &  0.1494 &  0.1497 &  0.1497 
&  0.1499\tablenotemark{c}&
 0.1498\tablenotemark{c}&  0.1502  & 0.1499 &
 0.1498& 0.1502& 0.1497 & 0.1497 &
 0.1497 & 0.1499  \\
FWHM\tablenotemark{d} & 1830 & 1150 & 1480 
& 1380\tablenotemark{c}
& 1380\tablenotemark{c} & 1380\tablenotemark{c}
& 1380 & 1380 & 1400 & 1390 & 1520
& 1340 & 1170\\
\enddata
\tablecomments{
$^{\rm a}$Normalized to 1.0. 
The observed flux of H$\beta$, in units of 
$10^{-15}$~erg~s$^{-1}$~cm$^{-2}$, is
1.2 (blue system), 3.2 (red system), and 6.0 (broad system).
$^{\rm b}$Sum of [S\,II] $\lambda6716$ and [S\,II] $\lambda6731$.
$^{\rm c}$Fixed. 
$^{\rm d}$In units of ${\rm km\,s^{-1}}$. Corrected for instrumental resolution.
}
\end{deluxetable*}


SDSS\,J1316+1753 is an emission-line galaxy at redshift $z = 0.15$, 
with an absolute magnitude $M_i=-21.9$ (calculated from 
the SDSS $i$\_psf magnitude).
The SDSS spectrum, corrected for Galactic extinction,
is displayed in Figure\,1(a), and a zoom on the H$\beta$-\oiii\
and H$\alpha$-\nii-\sii\ complex is shown in Figures\,1(b) and
1(c). Many emission lines are detected (Table\,1),
including the relatively rare feature of \ariii\,$\lambda$7136.

All emission lines show double-peaked profiles.
Among each double peak,
the emission-line system at lower redshift is hereafter
referred to as ``the blue system'', the system at higher
redshift as ``the red system''.
Close inspection of the strongest emission lines reveals the presence of
an underlying broad component, which is clearly detected in
\oiii\ $\lambda\lambda4959,5007$, \oii\ $\lambda3727$, \neiii\ $\lambda3869$
and in the Balmer lines H$\alpha$ and H$\beta$.

\begin{figure*}
\begin{center}
    \includegraphics[width=105mm]{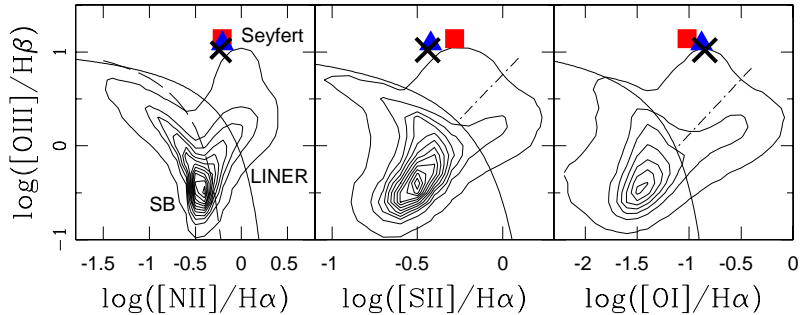}
\end{center}
\caption{
Location of the three emission-line systems in diagnostic diagrams.
The red square refers to measurements of the red system,
the blue triangle to the
blue system, and the black cross to the broad component.
The contours represent the distribution of SDSS narrow-line galaxies
of Kauffmann et al. (2003).
The single solid and dashed lines correspond to the dividing lines
between AGNs and star-forming galaxies of Kewley et al. (2001)
and Kauffmann et al. (2003), respectively.
The dot-dashed line marks the dividing line between Seyferts
and LINERs (Kewley et al. 2006).
All three emission-line systems are located in the Seyfert regime.
}
\end{figure*}

Emission lines were analyzed by fitting Gaussians, superposed on
a local continuum.
We use single Gaussian components to describe the narrow lines
of the blue and red system, and one extra Gaussian component
to model the broad system.
Fit parameters of all three Gaussians were the FWHM, flux,
and central wavelength.
Occasionally, for deblending or faint lines,
we fixed the FWHM and central wavelength in the fitting process (see Table\,1).
All quoted FWHMs were corrected for the instrumental resolution.
We have first focused on the brightest emission line,
\oiii\ $\lambda5007$.
The three-Gaussian parameterization describes the observed line profile well.
The blue and red system are
separated by $\sim$500\,${\rm km\,s^{-1}}$ in
velocity space.
The blue \oiii\ line at $z=0.1491$
is characterized by ${\rm FWHM=140\,km\,s^{-1}}$,
while the red line at $z=0.1510$ shows
${\rm FWHM=480\,km\,s^{-1}}$.
The broad component peaks at a redshift $z=0.1498$, intermediate
between the two narrow lines, and has ${\rm FWHM=1380\,km\,s^{-1}}$.
Gaussians with similar widths and redshifts also successfully fit H$\beta$
and other emission lines (see Table\,1).
The emission-line strengths and
ratios in the red and blue system are remarkably similar.
Emission-line ratios place both systems in the AGN regime
in diagnostic diagrams (Figure\,2); the same holds for the broad component.

No broad component is detected in the Balmer lines -- apart from the component
at FWHM ${\rm \sim 1400\,km\,s^{-1}}$ already mentioned, which is similar
to that seen in the other narrow lines, and therefore very likely has
the same origin, different from a classical broad-line region (BLR).
Faint absorption features from the host galaxy are visible in the spectrum,
and we identify absorption features from 
\nai\,D and \caii\,K.
Measurements of higher signal-to-noise ratio (S/N) 
follow-up optical spectra would allow us to
derive $\sigma_*$ directly from the absorption features.

We have checked the radio archives and found that SDSS\,J1316+1753
was detected in the NRAO VLA Sky Survey (NVSS) and the Faint Images of
the Radio Sky at Twenty-Centimeters (FIRST) survey at a flux level of
$S_{\nu,{\rm NVSS}} = 10.3$ mJy and $S_{\nu,{\rm FIRST}} = 11.4$ mJy,
respectively.

\section{Discussion}

SDSS J1316+1753 is exceptional in showing double-peaked
structure in all of its narrow emission lines
plus an underlying broad component in most forbidden lines, 
likely peaking in between the two narrow systems.
Given previous reports of spectroscopic binary AGN candidates
(with double-peaked \oiii, and perhaps H$\beta$), 
SDSS J1316+1753 suggests itself 
as another candidate which, in particular, has {\em all}
of its emission lines double-peaked.  
However, line splitting of individual or all emission lines
has also occasionally been observed before in single
AGNs (e.g., Heckman et al. 1981;  Veilleux 1991a, 1991b; 
Rodriguez-Ardila et al. 2006),
and in a few compact radio sources (e.g., Holt et al. 2008). 
It has been suggested to be
linked to the geometry of, or local physics in, the
NLR, or the influence of radio jets.
Therefore, other processes not related to binary AGNs can also
be imagined
which can explain the phenomenon of narrow double-peaked 
emitters. We discuss a large number of possibilities in turn. 
SDSS J1316+1753 with its rich, bright, 
and unusual emission-line spectrum allows us to test such models.

(1) {\it Superposition of two unrelated, but spatially close, AGNs
(a ``dual'' AGN).}
This possibility is very unlikely for two reasons.
First, the great similarity in the line fluxes and ratios, 
including the rare \ariii\ transition would
then be pure coincidence. Second, and more seriously,
the presence of a broad component which peaks at {\em intermediate} velocity
in between the red and blue system is inconsistent with
a chance projection.

(2) {\it Outflow.} 
A biconical outflow could naturally produce double-peaked 
emission lines. However, in order to affect the whole NLR, including 
the low-ionization lines like \sii, an unusually powerful outflow 
would be required. Furthermore, a powerful outflow would likely produce
a strong ionization and velocity stratification between the high-ionization 
lines that originate closer to the nucleus and the low-ionization
lines at larger distances (e.g., Komossa et al. 2008). 
While an ionization stratification is apparent in the
red system, there is none in the blue system.
 
(3) {\it Jet-cloud interaction.} 
The local interaction of radio
jets with NLR clouds is known to produce locally
very complex \oiii\ profiles including broad components in radio galaxies 
(e.g., Whittle et al. 1988; Morganti et al. 2007; Holt et al. 2008). 
However, these processes are sometimes very
localized, and do not usually dominate the integrated \oiii\ profile
from the whole NLR (see also the discussion
by Comerford et al. 2009a; but see Heckman et al. 1981; Veilleux 1991b).
Nevertheless, we explore such a scenario for SDSS J1316+1753.
Since no classical BLR is detected,
SDSS J1316+1753 is of type 2, which in
the context of the unified model of AGNs would imply an edge-on geometry with
radiation cones and jets more likely perpendicular to our line of sight.
A strictly perpendicular geometry would not produce double-peakedness, since
the velocity component in our line of sight would be small. However, we could
imagine intermediate viewing angles. Entraining the bulk of a classical NLR
in a two-sided jet is not easy to achieve, since jet-cloud interaction usually
produces cloud fragmentation rather than entrainment (see, e.g., the discussion
by Komossa et al. 2008). Strictly localized interaction of the jet
with a single cloud on each side of the nucleus might produce double-peaked
lines and perhaps shock ionization, 
but in order for this profile to dominate the observed emission lines,
the rest of the NLR would have to be weak or absent.
We have measured the \oiii\ flux in the
red system. Its luminosity of ${\rm 3\times 10^{42}\,erg\,s^{-1}}$  
is typical for bright Seyfert galaxies or quasars,
well above the emissivity expected
from just two single NLR clouds.
The high line luminosity we measure also argues against
a scenario, in which we only
observe the very inner NLR of SDSS J1316+1753
(line splitting in \oiii\
has occasionally been seen in the cores of AGNs;
e.g., Mazzalay et al. 2009), the rest of the NLR being absent.
Another variant of jet-induced line emission cannot be 
excluded. 
Young radio jets in gas-rich
mergers are not uncommon, and their
interaction with circumnuclear gas is
known to produce complex emission-line
profiles, including broad components 
(e.g., Holt et al. 2008; Inskip et al. 2008; 
see Morganti et al. 2009b for a review).
Fast autoionizing shocks (Dopita \& Sutherland 1996)
might then also produce local ionizing
continua which would be efficient in ionizing ambient gas, expanding
around the jets. 
 
{\it (4) Special NLR geometry.}
In purely geometrical terms, a spherical core component, plus a 
flat extended disk component seen from the side can naturally 
produce the red and blue narrow system together with the 
intermediate broad system. The broad component would arise in the 
spherical very inner part of the NLR (or outer BLR), the red 
and blue system in the disky part.
The broad component would then, in terms of its FWHM, 
correspond to the outer BLR. 
However, the conditions in the outer BLR are not favorable to produce
lines like \oii. \oii\ has low critical density, and is of relatively
low ionization, while in the outer BLR, high density, and a higher 
degree of ionization would generally prevail.
The red and blue system could plausibly arise in the classical NLR, 
if the NLR followed a strict disk geometry (e.g., Greene \& Ho 2005),
or followed the geometry of an inner bar or an inner two-sided spiral. 
An attractive feature of this scenario is that it is consistent with
the great similarity of the line ratios in the red and the blue system
(including the detection of \ariii\ in both of them),
because only a single ionizing continuum is needed,
and because the physical condition in the NLR on both sides would be expected
to be similar (in terms of metal abundances, cloud densities, and column 
densities). 

If emission-line ratios are dominated by photoionization,
we can use the ratio of \oii/\oiii\ to estimate the ionization
parameter (Komossa \& Schulz 1997). 
We find log $U \simeq -2$ for the red and blue system,
assuming that the density we measure from the \sii\ 
ratio ($n_{\rm e} \approx 30$ cm$^{-3}$ for the blue system 
and $n_{\rm e} \approx 400$ cm$^{-3}$ for the red system) reflects
the average density of the system and that there are no large density
inhomogeneities, and without performing any extinction correction. 
The measured total H$\beta$ luminosity implies
a minimum rate $Q$ of hydrogen-ionizing photons of 
$Q = 2.7 \times 10^{54}$ s$^{-1}$
(assuming unity covering fraction of the H$\beta$ emitting clouds),
which then translates into a {\em lower limit} on the 
distance of the NLR of 0.4\,kpc
of the red system and 1.6\,kpc of the blue system.   

{\it (5) A galaxy merger with one (or two) active nuclei.}
In this scenario, SDSS J1316+1753 would consist of 
two galaxies with two separate
NLRs in the process of merging.
The SDSS image of SDSS J1316+1753 indeed indicates a possibly distorted
morphology and an excess of companion galaxies. 
{\footnote[3]{The velocity between the red and
blue system ($\sim$400--500 \kms) is rather high, above that typically
expected for a bound merger. Previously reported double-peaked emitters
had peak separations of 730 \kms and 630 \kms, respectively
(Zhou et al. 2004; Gerke et al. 2007).
Systematic shifts of single-peaked emission lines
with respect to host absorption lines observed by Comerford et al. (2009a) are
on the order of 100--300 \kms.}}
In a merger scenario, we can have either two (obscured) 
accreting black holes each
illuminating its own NLR (the scenario was favored
previously to explain some \oiii\ double-peaked emitters),
or else a single AGN illuminating the interstellar media
of both galaxies, i.e., both NLRs. 
The similarity
in line fluxes and ratios of the red and blue system argues
for a relatively small separation of the two NLRs, 
no matter whether both of them see the ionizing continua
of one, or of two AGNs.  
Receiving similar ionizing continua
would potentially  
make line ratios more similar.

In the merger scenario, we expect that the host galaxy absorption features
should more closely match the redshift of the emission lines from the more
massive host galaxy  (while in jet/outflow-related scenarios the host redshift
should be in between the two emission-line redshifts). We do 
detect faint absorption features from the host galaxy in the spectrum
of SDSS J1316+1753. 
By comparing the observed peak location of the two absorption features
with the redshifts of the red and blue system, we
tentatively find that they are either more consistent with the red system,
or are located in between the two emission systems.
In the context of the merger scenario, the {\em broad} component of the 
forbidden lines may
arise in gas that is related to starburst-driven superwinds. 
Mergers frequently trigger enhanced starburst activity, and the 
well-known merger and binary AGN NGC 6240 does show off-nuclear 
broad Balmer lines which are likely linked to superwinds (Heckman et al. 1990).
If the two narrow \oiii\ systems represented two classical NLRs of the two
merging galaxies, we could use their widths as a proxy for gaseous velocity
dispersion 
(e.g., Nelson 2000). 
Actually doing so, and assuming the 
$M_{\rm BH} - \sigma$ relation to
apply (Ferrarese \& Ford 2005), we obtain black hole masses of 
$2 \times 10^8$\msun\ and $5 \times 10^5$\msun;
the second value is very low 
and would then imply a minor merger. 

\section{Future observations}
A number of future observations can distinguish between
different scenarios for SDSS J1316+1753 and double-peaked
narrow-line emitters in general. These would either focus on follow-ups
of SDSS J1316+1753, or would address class properties of larger
samples systematically selected from SDSS.
{\it Hubble Space Telescope} imaging and 
Space Telescope Imaging Spectrograph spectroscopy 
will reveal the detailed morphology of
the host galaxy of SDSS J1316+1753
and will determine or constrain the extents and loci
of the red, blue, and broad intermediate system.
High-resolution radio imaging might reveal the presence of jets,
and/or provide evidence for two active nuclei,
if present.
While detailed emission-line analysis and modeling of individual
objects likely 
holds the key to understanding the physical processes involved,
a systematic search for similar objects in the
SDSS data base (H. Lu et al. 2010, in preparation) 
to study the bulk properties
of double-peaked emitters is important, too.
Measurements of the host absorption
features in spectra of high S/N will show whether the absorption features
coincide in redshift with one of the 
two narrow emission-line systems, 
or are always
located in between.  The question of whether one AGN in a merger scenario
is sufficient to illuminate both NLRs, or two AGNs are required, 
can be addressed by searching for ionization stratifications
in the narrow lines, in cases where high-resolution
imaging is not available. If only one AGN is illuminating
two NLRs, the second NLR should lack an ionization
stratification.
Of great interest will be the frequency of
double-peaked emitters in type 2 versus type 1 AGNs, and in
high-luminosity AGNs (quasars)
versus low-luminosity systems (Seyfert galaxies).
If they are more abundant in type 2s, that would hint at geometrical effects
playing a role. If they are more abundant in quasars than in Seyferts, that 
would be consistent with
a merger scenario, since quasars are likely powered by major mergers.

In summary, a galaxy merger with two (or one) AGNs
illuminating the NLRs of the two merging galaxies is
one possible scenario to explain narrow double-peaked emitters,
especially SDSS J1316+1753 because not just one
but all emission lines are double-peaked, and an additional
broad component in forbidden lines is present at
intermediate velocity. 
However, the line separation of the red and blue narrow system 
of about 400--500 \kms\ is relatively 
large for a bound merger; and the
remarkable similarity in line fluxes and ratios of the red and blue system
needs to be explained.
While localized jet-cloud interaction in an otherwise gas-poor
NLR is unlikely because of the high observed emissivity in the emission lines,
newly triggered jets in a gas-rich merger,
and special NLR geometries remain plausible possibilities.
Confirming or rejecting the binary AGN hypothesis is of great
interest in the context of studying BH mergers, mechanisms of
fueling and gas transport into the galaxy core, and/or to learn more
about NLR dynamics and geometry.
SDSS J1316+1753 
might be a Rosetta stone 
in this regard, due to its rich and very bright narrow-line spectrum
and the presence of the remarkable broad component in the forbidden lines.\\

{\em Note added after submission.}
After submission of our manuscript,
two more papers on double-peaked emitters selected from the SDSS 
appeared in preprint form 
(Smith et al. 2009; and Liu et al. 2009). 
These two, and a preprint by Wang et al. which we received, 
are complementary to ours
in that they study sample properties of narrow double-peaked emitters
in type 1 and type 2 AGNs, while our work focuses in detail on the
emission-line properties of a single object. 

\acknowledgments
We thank M. Gaskell, B. Gerke, C. Li, F. Liu, H. Lu, 
D. Merritt, J. Wang, H. Zhou, and the participants
of the KIAA workshop on ``Massive Black Hole Binaries and Their
Coalescence in Galactic Nuclei'' for useful discussions. 
This work was supported by NSFC\,10873017,
the 973 program (2009CB824800), and the MPG-CAS exchange program.

\end{document}